%% file: main.tex
\documentclass[a4paper,11pt]{article}
\pdfoutput=1 

\usepackage{jheppub} 

\usepackage[T1]{fontenc} 

\title{\boldmath Quasi Anomalous Knowledge: Searching for new physics with embedded knowledge }

\arxivnumber{2011.03550}

\author[a,b]{Sang Eon Park}
\author[a,b]{Dylan Rankin}
\author[a,b]{Silviu-Marian Udrescu}
\author[a]{Mikaeel Yunus}
\author[a,b]{Philip Harris}

\affiliation[a]{Laboratory for Nuclear Science, Massachusetts Institute of Technology,\\ 77 Massachusetts Ave, Cambridge, MA 02139, U.S.A.}
\affiliation[b]{The NSF AI Institute for Artificial Intelligence and Fundamental Interactions}

\emailAdd{sangeon@mit.edu,drankin@mit.edu,sudrescu@mit.edu,\\myunus@mit.edu,pcharris@mit.edu}

\abstract{
Discoveries of new phenomena often involve a dedicated search for a hypothetical physics signature. Recently, novel deep learning techniques have emerged for anomaly detection in the absence of a signal prior. However, by ignoring signal priors, the sensitivity of these approaches is significantly reduced. We present a new strategy dubbed Quasi Anomalous Knowledge (QUAK), whereby we introduce alternative signal priors that capture some of the salient features of new physics signatures, allowing for the recovery of sensitivity even when the alternative signal is incorrect. This approach can be applied to a broad range of physics models and neural network architectures. In this paper, we apply QUAK to anomaly detection of new physics events at the CERN Large Hadron Collider utilizing variational autoencoders with normalizing flow.
}

\begin{document} 
\maketitle
\flushbottom

\nocite{Kasieczka:2021xcg, Bortolato:2021zic, Stein:2020rou, Dillon:2020quc, Mikuni:2020wpr, Collins:2018epr, Collins:2019jip, Collins:2021nxn, Andreassen:2020nkr, Benkendorfer:2020gek}

\section{Introduction}
\label{sec:intro}
\input{introduction.tex}

\section{Concept}
\label{sec:concept}

\input{concept.tex}

\section{Normalizing Flows}
\label{sec:normalizing_flow}
\input{normalizing_flow.tex}

\section{Results}
\label{sec:results}
\input{results.tex}

\section{Conclusions and outlook}
\label{sec:conclusion}
\input{conclusion.tex}

\section{Acknowledgements}
\label{sec:acknowledgment}
\input{acknowledgements.tex}


\bibliographystyle{JHEP}
\bibliography{references}

\end{document}

%% file: introduction.tex
With no evidence for new physical phenomena, many physicists at the CERN Large Hadron Collider (LHC) are asking themselves a critical question: \emph{Am I searching for new physics in the right way?} Despite ten years of exhaustive research at the LHC, a rapidly growing cohort is becoming concerned that we have somehow missed a new fundamental physics discovery. Within particle physics and beyond, the identification of new physical phenomena has often been unexpected. With advances in deep learning (DL), a series of new approaches can improve the search for anomalous signatures. In this paper, we will present a new deep-learning-based anomaly search algorithm. This algorithm is broadly applicable to many fields of physics. 

Recent DL-based anomaly detection within high energy physics has largely focused on searching for anomalous signatures in the complete absence of a signal prior. In this scenario, two fundamental approaches have been considered: 
\begin{itemize}
\item Isolate two distinct datasets that contain signal and background with different proportions, then try to find a deviation between them.~\cite{Metodiev_2017,Collins_2018,Collins_2019,Nachman_2020}
\item Embed our knowledge of known physics processes into simulation or a DL algorithm, such as an autoencoder,  and then look for events with a low likelihood of being a known physics process.~\cite{Heimel:2018mkt,Farina:2018fyg,Cerri_2019,Kuusela_2012,roy2020robust,Heimel_2019,roy2020robust,Blance_2019,Hajer_2020,dagnolo2020learning,D_Agnolo_2019,Romao:2020ocr} 
\end{itemize}
In the first approach, colloquially referred to as classification without labels (CWoLA), conventional discrimination algorithms are used to separate the two datasets\cite{Metodiev_2017,Collins_2018,Collins_2019,Nachman_2020}. Care must be taken to ensure that selection biases are mitigated so that the only discernible difference within the discrimination algorithm is the presence of an unknown physics signal. The second approach attempts to embed into a likelihood discriminant a complete knowledge of physics processes within a selected region. An excess of events with low likelihood constitutes a new physics signature. This second method broadly comprises models that utilize deep learning autoencoders. However, when using large, high dimensional datasets, complete knowledge of all expected physical processes can become quite complicated. It can lead to reduced sensitivity\cite{Heimel:2018mkt,Farina:2018fyg,Cerri_2019,Kuusela_2012,roy2020robust,Heimel_2019,roy2020robust,Blance_2019,Hajer_2020,dagnolo2020learning,D_Agnolo_2019}. Recently, hybrid approaches have started to emerge, which aim to utilize aspects of both methods\cite{amram2020tag}. 

When comparing the two approaches, the CWoLA approach is often more sensitive, provided a signal-enriched region is present\cite{Nachman_2020}. This enlarged sensitivity results from the implicit assumption on the signal properties; the signal is localized within a specific kinematic region. In other words, CWoLA assumes that we can find a signal enriched region. Signal priors frequently lead to enhanced sensitivity since they minimize the possibilities that must be explored when attempting to search for an anomaly. For many new physics models within HEP, several fundamental assumptions can be applied to all potential signals without loss in generality. However, these assumptions are often not embedded with a neural network aimed at anomaly detection. Thus, the network cannot infer whether an observed anomaly within the data violates fundamental symmetries of nature required for a new physics model. For example, when a massive particle decays, its decay products fall within a cone determined by the particle's energy and Lorentz invariance. When a generic DL algorithm attempts to probe data, it has no knowledge of Lorentz invariance\cite{Butter_2018,choy20194d,Bogatskiy:2020tje}. 

By relying on \emph{one} anomaly metric that measures any deviation, whether it be physical or not, we may miss the chance to apply fundamental physical laws about how new physics may appear, thus wasting our prior physics knowledge. If we can incorporate prior knowledge into the search, it should be possible to either improve the search's sensitivity or, at worst, restrict the network complexity. Within physics, several ideas have emerged to incorporate physical symmetries into the neural network. These ideas involve modifications to the network architecture so that networks implicitly respect physical laws. In this paper, we consider an alternative approach. Instead of modifying the network, we rely on extending the anomaly algorithm by exploiting a second self-learned space explicitly targeting a class of new physics signatures, i.e. an added signal prior. This space embeds the most critical symmetries respected by new physics signature through self-learning, thus allowing for detection of all anomalies that are broadly similar to the desired signal. 

With our modified anomaly detection algorithm, we extend the use of signal priors to anomaly searches by developing a mechanism to add signal priors without degrading the sensitivity of a pre-existing model-independent search. Through our approach, signal priors, which may or may not be accurate signal descriptions, can be embedded within an anomaly search. Since priors are systematically added to construct information, we refer to this technique as Quasi-Anomalous Knowledge, or simply QUAK.

\begin{figure}[htbp]
\centering
\includegraphics[width=.95\linewidth]{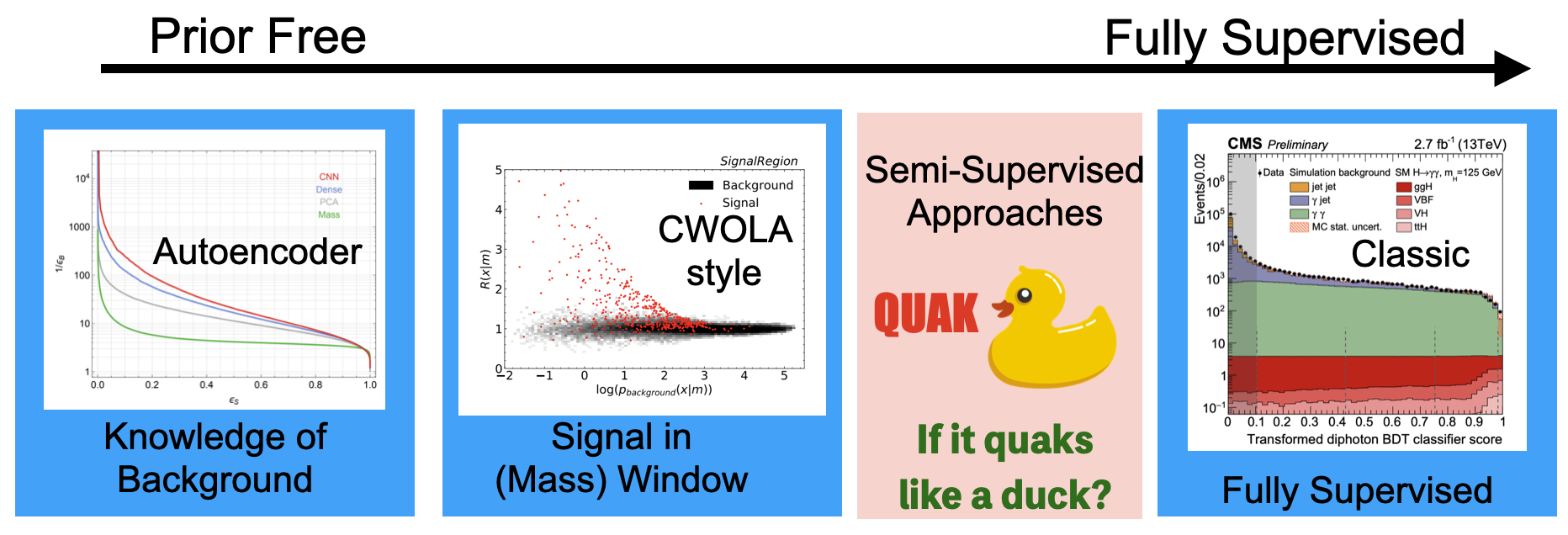}
\caption{Overview of Deep Learning based methods used in high energy physics. The prior here indicates the amount of a signal prior used within the search. This paper presents QUAK, which falls into a region labeled by the red shaded area. }
\label{fig:algos}
\end{figure}

To understand the context of QUAK within the scope of search strategies at the LHC, consider Figure~\ref{fig:algos}. Currently, most searches for new physics at the LHC involve the search for a specific, well-motivated physics signature (rightmost region of Figure~\ref{fig:algos}). This is performed by considering a specific signal prior and optimizing selection towards this signal prior. A dedicated analysis is performed with each individual search to account for mismodeling and systematic effects. 

The critical aspect of this type of search is that one constructs the selection, discriminator, and sensitivity estimate under a prior for what the signal model is assumed look like. As an example, consider a search for black holes. First, we hypothesize what a black hole signature would look like. From our hypothesis, we construct a Monte Carlo simulation of black hole signal events. We proceed to build an analysis around this assumption. Barring a discovery, our final search will give us bounds on the production of black hole processes under these assumptions. It may not give us a bound on any other process and it might not even cover all possible black hole signatures. However, it will have shown that, in the region of collision data where events are black hole like, we observe good modeling of predicted background processes with the data. While the choice of a specific signal signature is restrictive, it has the benefit that, for physics models that predict similar signatures to our signal, we can make powerful constraints. 

With QUAK, we aim to embed this choice of a signal prior within a more generic search for an anomaly. Anomaly detection algorithms typically forego the signal prior, and, as a consequence, rely on some metric to determine what an anomaly is. With QUAK, we aim to create a space that allows us to interpolate from a dedicated search with a clear signal prior to a prior-free search that quantifies physical anomalies and implicitly emphasizes signal-like features. As a consequence, QUAK can be considered a semi-supervised approach.

In the following sections, we will introduce the QUAK algorithm. 
First, to show the generality of this approach, we demonstrate the usage of QUAK on the MNIST dataset\cite{mnist}. 
Then, we present this work in the context of the LHC Olympics 2020 anomalous dataset\cite{kasieczka_gregor_2019_3596919}.

%% file: concept.tex
\begin{figure}[htbp]
\centering
\includegraphics[width=.95\linewidth]{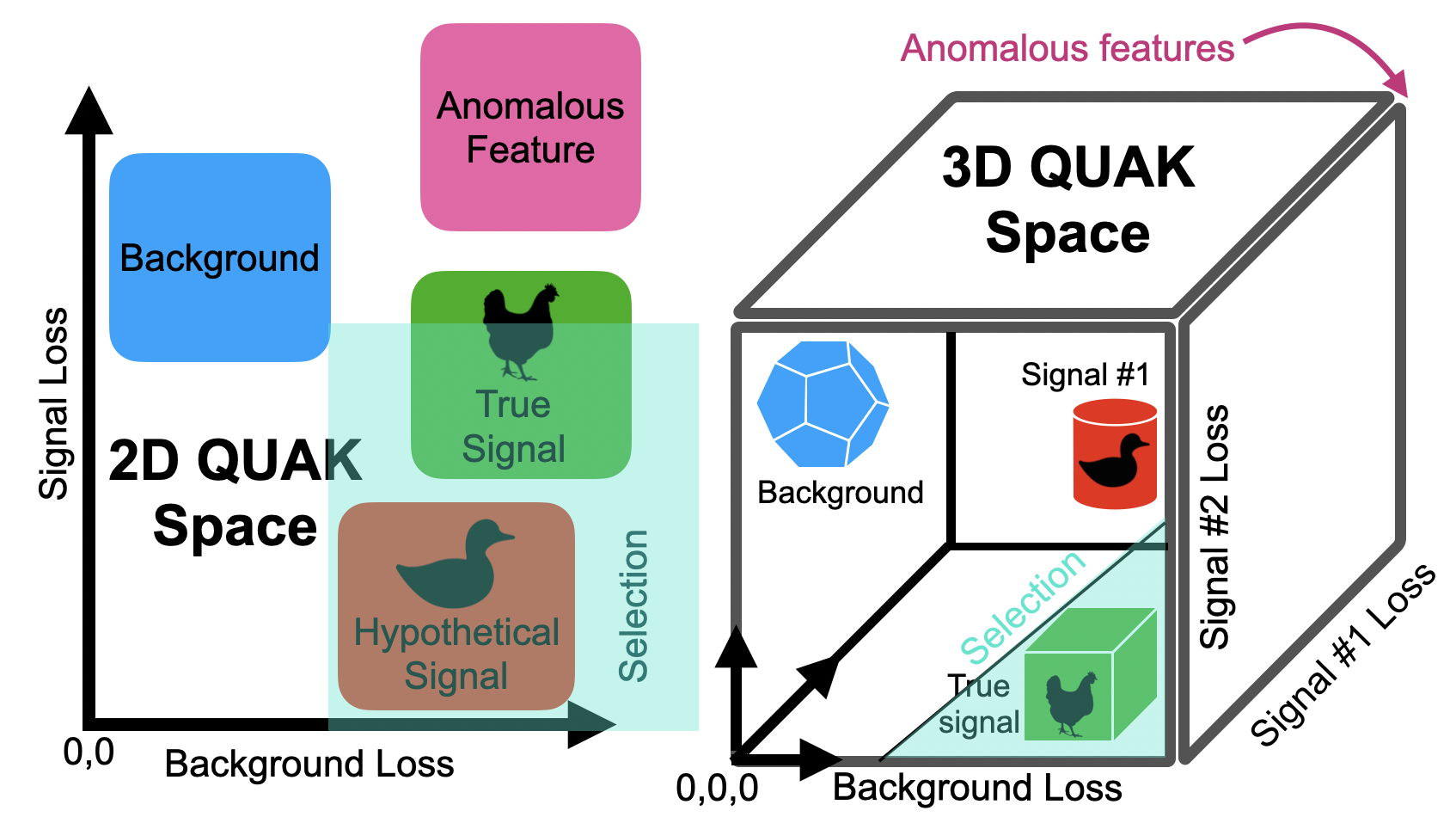}
\caption{Illustration of the 2D QUAK space construction (Left) and the 3D QUAK space construction (Right). In each space, we list where background events would roughly be located (blue), signal events would roughly be located (red and green), and anomalous unphysical features would be located(magenta). We split the signal events between hypothetical, guessed signal, used to construct the space (red), and the location of a potential true signal(green). The selected region (cyan) indicates roughly where anomalous new physics signals are most likely to appear.}
\label{fig:quak}
\end{figure}

New models have emerged within the deep learning community that utilize semi-supervised learning to construct a physical, self-organized space \cite{chen2020big,ouali2020overview}. Semi-supervision applied to deep learning works by training on both labeled and unlabeled data. With the unlabeled data, an unsupervised network is trained (autoencoder). With the labeled data, a supervised classification is applied using the unsupervised network's intermediate latent space. The trainings are often done simultaneously. Other approaches can be considered semi-supervised learning, such as using analytic continuous to fill poorly understood spacer. 

By performing the unsupervised training, the network constructs a latent space of self-organized patterns. By performing the supervised training, we label regions within the latent space with the labeled data. Consequently, objects present in the unlabeled data but not in the labeled data will be labeled by a superposition of its nearest labeled identifiers.  Additionally, the construction of latent space from the unsupervised training using a variational approach by sampling Gaussians within the latent space can ensure that the latent space is continuous.

Semi-supervised constructions represent a different approach to training neural networks when compared with supervised and unsupervised. Semi-supervised networks are very robust to variations in the data, and, in some cases, these networks have been found to exceed the performance of supervised networks\cite{hendrycks2019using}. Within the context of anomaly detection, semi-supervision has been found to be effective for anomaly detection\cite{ruff2020deep,hendrycks2019deep}, even very recently within physics\cite{cheng2020variational}. 

QUAK builds on the concept of semi-supervision. However, the algorithm differs from other semi-supervised approaches in that we rely more on unsupervised networks when constructing our algorithm. To construct QUAK we: 
\begin{itemize}
 \item choose a set of N samples that reflect the anomaly search; this is typically a background and a set of signals, with the signals that capture the physical features of a potential new physics signature,
 \item train a separate unsupervised networks on each signal or background sample, yielding N unsupervised networks,
 \item construct an N-dimensional "QUAK" space consisting of the loss for each unsupervised network, and
 \item search within bins of QUAK space for anomalous signals. 
\end{itemize}
The construction is semi-supervised in that we use the signal priors as labels for QUAK space construction.
Figure~\ref{fig:quak} illustrates the concept of QUAK space. In each axis, we plot the loss of the unsupervised network trained on a specific sample. For a 1-dimensional QUAK where we train just on the background sample, QUAK reduces to a typical autoencoder based anomaly search. However, the advantage of QUAK is that the added signal dimensions help to disambiguate anomalies that are signal like from anomalies that are just anomalous, such as an anomalous event resulting from a detector glitch. 

Within QUAK space, a background event will have low background loss and high signal loss. In the 2D QUAK plot, this corresponds to the top left of the region. An anomalous event with features similar to the chosen signal will have a low signal loss and large background loss. This will occur in the bottom right region of 2D QUAK space. An anomalous event different from both the chosen signal, and the background will have large signal and background loss. This region will include events coming from detector glitches. Extending this beyond 2 dimensions, we can add additional, different, signals. These additional signals can further improve the anomaly search since they help to disambiguate anomalies with features similar to the chosen signal. By searching in the region of large background loss and low signal loss, on all signal dimensions, we isolate a region of anomalies that have physical, signal-like features. This space greatly enhances the ability to find new physics signatures provided they have a sufficient set of features to be captured by the anomaly algorithm. 

For many classes of new physics models at the LHC, there is a broad set of underlying physics features that can be assumed about any new signal (e.g. Lorentz invariance, lack of QCD color-flow with the event). These assumptions can restrict an anomaly search by highlighting, at low loss, features characteristic of a signal. With QUAK, we effectively embed these assumptions into our search while preserving much of the model-independence of the investigation. 

For the network architecture of the unsupervised networks in QUAK, we utilize variational auto-encoders (VAEs) and normalizing flow VAEs. VAEs have been found to give a continuous latent space, which allows for the capture of physical properties within the latent space. Normalizing flows allow for the latent space to be irregular and not necessarily Gaussian. While the exact deep learning architecture is not critical, we focus on normalizing flow since it has been shown as a powerful tool for representing physical models\cite{rezende2020normalizing,Albergo_2019,Kanwar_2020,brehmer2020flows,bothmann2020,gao2020,iflow2020,Nachman_2020,choi2020,lu2020,bieringer2020,hollingsworth2021}. In the QUAK construction, our signal choices can be thought of as "approximate-priors" since they help direct the space of searches towards signals with similar features. This, in turn, leads to a potential model dependence in the anomaly search. However, as we will see with QUAK, the signal choice can significantly differ from the observed anomaly and give an enhanced sensitivity compared with other signal-less anomaly search algorithms.

Finally, we note that while we consider QUAK to be a semi-supervised network, QUAK deviates from other semi-supervised networks. We do not exploit a common latent space between the supervised and unsupervised component of the network.

%% file: normalizing_flow.tex
To construct QUAK, we rely on normalizing-flow variational autoencoders (NF-VAEs). A variational autoencoder (VAE) is, in essence, an autoencoder that samples from a multidimensional Gaussian distribution in the first layer of the latent space. The loss is computed by constructing the cross-entropy between the input dataset and the output dataset. An additional Kullback-Leibler (KL) divergence term is added to the loss to encourage the VAE to approximate the posterior with a multidimensional Gaussian distribution\cite{kingma2014autoencoding}. 

VAEs have several limitations. With a VAE, we assume that the latent space can be approximated using a series of compounded linear transformations of a multidimensional Gaussian distribution. Consequently, VAEs work best when the input variables are approximately Gaussian. If the input variables adhere to a distribution that is clearly skewed or multimodal, we require a large decoder to transform the Gaussian latent vectors to vectors that can accurately reconstruct the input data. If the decoder is too large, the latent space will be rendered useless. This phenomenon is known as the ``posterior collapse''\cite{bowman-etal-2016-generating}. 




To avoid the limitations of VAEs, one can use normalizing flows to transform a posterior distribution into a much more flexible distribution that is representative of the corresponding data ~\cite{rezende2015variational}. 
We incorporate normalizing flows into our VAEs by using them in our latent space. Specifically, if $z_0$ is the latent vector obtained via Gaussian sampling, and $z_1, \cdots, z_k$ are the latent vectors that follow, a series of normalizing flows $f_1, \cdots, f_k$ will translate the posterior of the latent space as follows:

\begin{align}
  z_k = f_k \circ \cdots \circ f_1 (z_0)
\end{align}

Furthermore, under the assumption that $z_i$ belongs to a distribution $q_i(z_i)$ for all $i \in \{0, \cdots, k\}$, we can write

\begin{align} \label {ladj}
  \log{q_k}(z_k) = \log{q_0}(z_0) - \sum_{i=1}^{k}\log{\left|\det\left(\frac{\partial f_i}{\partial z_{i-1}}\right)\right|}
\end{align}

Consequently, we have effectively transformed our space from the assumed Gaussian space utilized within the VAE. By transforming our Gaussian posterior (i.e. the latent space) into a more complex posterior, we can accurately account for jet variables whose underlying distributions are not necessarily Gaussian. This removes the need for a larger decoder, which inherently prevents a ``posterior collapse'' from occurring. Furthermore, we can discern from Equation \ref{ladj} that we must update our loss function to account for the introduction of a more expressive posterior $z_k$ . This we perform by modifying the loss as follows: 

\begin{align}
  \mathcal{L} = \mathcal{L}_{\text{reco}} + \mathcal{D}_{KL} - \sum_{i=1}^{k}\log{\left|\det\left(\frac{\partial f_i}{\partial z_{i-1}}\right)\right|},
\end{align}
where $\mathcal{L}_{\text{reco}}$ corresponds to the conventional autoencoder loss, and $\mathcal{D}_{KL}$ corresponds to the KL divergence term. Our new loss function applies to a much wider range of posterior distributions. In order to determine the optimal normalizing flow algorithm, a scan of various normalizing flow algorithms was performed. With each, we required that $\det\left(\frac{\partial f_i}{\partial z_{i-1}}\right)$ be computed in linear time, along with the requirement that $f_1, \cdots, f_k$ be invertible. For this scan, we considered planar flows, radial flows, non-volume preserving flows, and masked autoregressive flow. We found that masked autoregressive flow gave the best performance for anomaly search. With masked autoregressive flow, we have that 
\begin{align}
  f(x_{i}) = u_{i}\exp f_{\alpha_{i}}\left(z_{1-i-1}\right) + f_{\mu_{i}}\left(z_{1:i-1}\right) \rm{ with,} \\
  \left|\det\left(\frac{\partial f}{\partial z}\right)\right| = \exp\left(-\sum_{i}f_{\alpha_{i}}\right)
\end{align}
where $u_{i}$ is a randomly sampled number over a Gaussian distribution of width 1, and the functions $f_{\alpha}$ and $f_{\mu}$ are applied on the previous 1 through i-th values of the input latent space. In this approach, the series of iterations taking the previous inputs $z_{i}$ are used to progressively improve the ability of the model to construct a probabilistic distribution of the internal space used within the autoencoder. The choice of generating function, and the auto-regressive construction enable a broad range of distributions. Furthermore, the masking provides a scheme for efficient sampling and generation of events. 

More advanced normalizing flow techniques have recently been developed by various researchers for use within physics\cite{rezende2020normalizing,Boyda:2020hsi,brehmer2020flows,Albergo_2019}. Exploration of these models is an interesting and exciting avenue for future work.

%% file: results.tex
To demonstrate the effectiveness of QUAK. We present two examples of how it can be applied to anomaly searches. First as a demonstration, we apply QUAK on a test dataset constructed from the MNIST dataset\cite{mnist}. Then, we consider applying QUAK to the LHC Olympics 2020 dataset to find anomalous dijet signals\cite{kasieczka_gregor_2019_3596919}. 

\subsection{MNIST}

As a first example and to show that QUAK approach can broadly apply to many scenarios, we demonstrate the usage of QUAK on the MNIST dataset\cite{mnist}. The MNIST dataset consists of a set of images of handwritten single digits ranging from 0 to 9, with explicit labels for each of the digits corresponding to their actual number. Standard VAEs without a normalizing flow layer are very effective at describing the MNIST dataset. 

To emulate the hypothetical possibility of discovery, we split the MNIST dataset into ``known'' and ``unknown'' digits. For this example, we will choose the digit 9 as unknown. The dataset is split so that the digits 0 to 8 are known, but the digit 9 is not present anywhere in the training datasets. We then attempt to discover this digit by constructing a QUAK space classifier. 

We construct QUAK space on MNIST through a variational autoencoder~\cite{kingma2014autoencoding} with three dense layers on either end; we do not use a normalizing flow layer since we found it was not needed. The training for the VAE uses a loss function consisting of a binary cross-entropy of the input image with the output image. A KL divergence term with equal weight to that of the cross-entropy loss term is added to the loss function to constrain the sampled Gaussian means and widths to be close to unity. The QUAK loss is obtained by removing the KL divergence term. Figure~\ref{fig:archs} shows the deep learning architecture used. 

\subsubsection{Separation of digits with QUAK}

To test QUAK on MNIST, we separately train a dedicated VAE for each of N known digits, yielding an N-dimensional QUAK space ranging from the digits 0 to 8. Within this space, we try to find 9 by choosing a proxy signal, 7, which is similar to the true signal digit. The choice of 7 is intended to be an educated guess for how to find a signal. If we did not know the digit beforehand, we could systematically scan each digit, 0--8, and use it as a proxy signal to search for an anomaly. Additionally, the use of a proxy signal allows us to deal with high dimensional QUAK spaces since we can compress an N-dimensional QUAK space into a single dimension through the training of an additional dense network aimed at separating the proxy signal from the background. 

\begin{figure}[htbp]
\centering
\includegraphics[width=.35\linewidth]{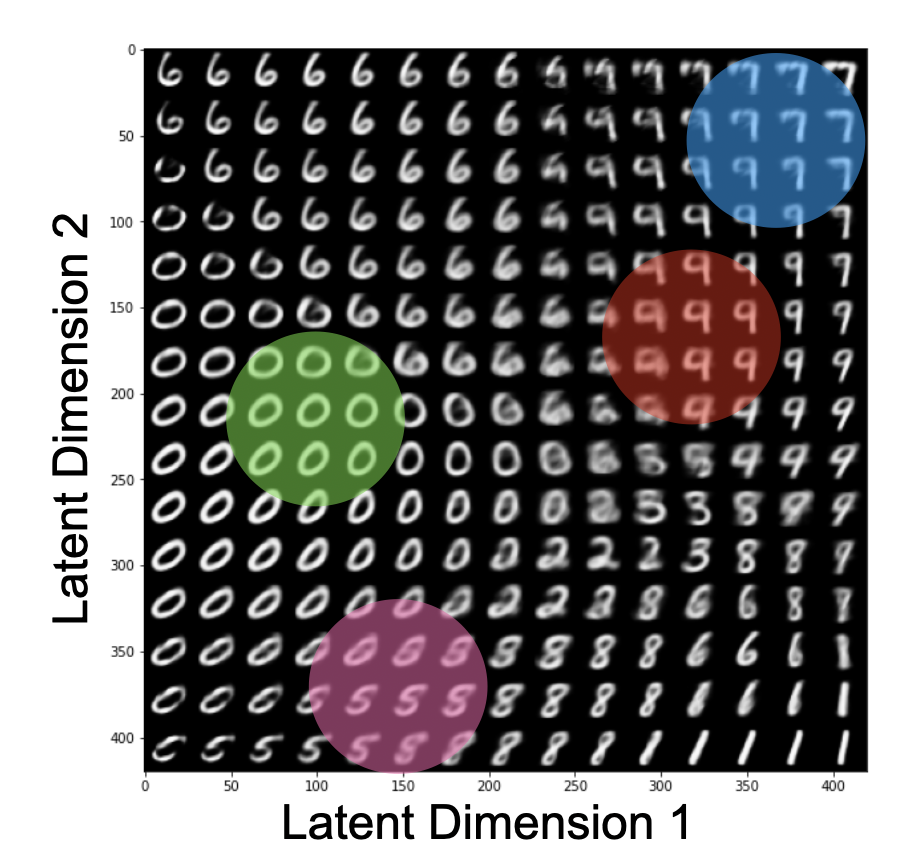}
\includegraphics[width=.45\linewidth]{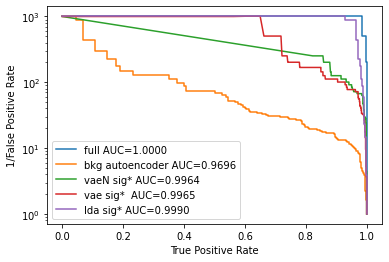}
\caption{ (Left) Illustrative plot of a 2-Dimensional latent space generated from a VAE trained on all of the digits 0-9. Colors indicate rough regions of interest for the digit 0 (green), 9 (red), 5 (purple), and 7 (blue). (Right) Exploration of different QUAK approaches as presented using the ROC applied on the separating of the digit 5 vs. the digit 9. We show the performance for (orange) a single autoencoder loss trained on the digit 5, (blue) a fully supervised network training the digit 5 vs. the digit 9, (red) a dense network trained to separate the digit 5 from 7 on a 2D QUAK space consisting of the loss of the digit 5 on one axis, and the loss of the digit 7 on another axis, (green) a dense network trained to separate the digit 5 from 7 on a 9D QUAK space consisting of the loss of each digit 0--8, (purple) a linear discriminant trained to separate the digit 5 from 7 on the same 9D QUAK space consisting of the loss of each digit 0--8.
}
\label{fig:mnist1}
\end{figure}

Figure~\ref{fig:mnist1} demonstrates the performance of various QUAK spaces in the separation of the digit 5 from the digit 9. In the language of physics searches, the digit 5 is the background and the digit 9 is the signal. We consider 3 different QUAK approaches:
\begin{itemize}
\item Scan of 2D QUAK space using 5 for $L_{bkg}$ and 7 for $L_{sig}$ and training a 3 layer dense network on the 2D QUAK space with 7 as the target signal and 5 as the background
\item Scan of the 9D QUAK space using the VAE loss of all digits 0--8, and training a 3 layer dense network on the 9D QUAK space with 7 as the target signal and 5 as the background
\item Scan of the 9D QUAK space using the VAE loss on all digits 0--8, and transforming the 9D space to a 1D discriminator using a linear discriminate trained on the 9D QUAK space with 7 as the target signal and 5 as the background
\end{itemize}
We contrast these approaches using a single autoencoder and a fully supervised network both trained to separate the digit 5 from 9. 

We find near optimal performance when using a linear discriminant with a proxy signal to separate signal and background. We observe the QUAK methods approaches the fully supervised network performance as the dimension of QUAK space is increased. When compared with a signal autoencoder, we again observe a rejection factor that is many orders of magnitude better for the same signal efficiency (true positive rate). 

\subsubsection{Proxy signal choice}
The choice of the proxy digit 7 to discriminate 9 is particularly appealing since both digits are quite similar. However, there is no guarantee that such an appealing choice would be present when searching for an anomaly. To illustrate the flexibility in proxy signal choice, we consider the instance where we use this same 7 digit proxy and dilute it within another less similar digit, 0. Following the diagram in the left of Figure~\ref{fig:mnist1}, the digit 0 is equally distant between the background digit 5 and signal digit 9 within a latent space generated on all digits. Figure~\ref{fig:mnist2} shows the performance in separating the 5 digit from the 9 digit with the diluted proxy. We again observe enhanced discrimination approaching the fully supervised limit as we inject more of the digit 7. 

\subsubsection{Latent vs Loss Space}
In both the dijet example presented above and MNIST, we have chosen to construct QUAK from the losses of dedicated (NF)VAEs. Another approach is to not use the VAE loss and, instead, using a latent space generated from a single VAE trained on both signal and background priors together. This approach reduces the number of trainings from N-signal or background priors to a single training. However, it requires a significantly larger and more flexible network capable of capturing all different features. 

\begin{figure}[htbp]
\centering
\includegraphics[width=.45\linewidth]{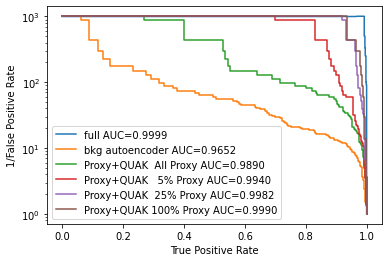}
\includegraphics[width=.45\linewidth]{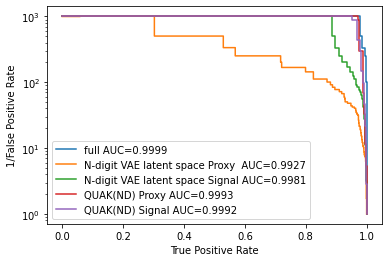}
\caption{
Performance in separating the digit 5 from 9 for a fully supervised network (full) with (Left) a linear discriminant on the space of digits 0-8 using a proxy signal that is X\% digit 7 and 100-X\% digit 0. (Right) Performance comparison of QUAK compared to an 9-dimensional latent space from a VAE trained on all digits 0--8 to separate out the digit 5 with the digit 9  with QUAK. To perform discrimination on the latent space a supervised network is trained on the latent space produced from the digit 5 against a latent space produced from the digits Signal(9)/Proxy(7). Performances are compared to QUAK space using a linear discriminant on the 0--8 digit loss space again using either Signal(9)/Proxy(7). }
\label{fig:mnist2}
\end{figure}

In Figure~\ref{fig:mnist2}, we compare the performance of a VAE with a 9-dimensional latent space trained on all 9-digits ranging from 0--8 simultaneously. We then use the latent space's N-dimensional output to train a three-layer multilayer perceptron to separate the digit 5 from 9 (signal) or 7 (proxy). We contrast this approach with the QUAK construction, whereby we train a linear discriminator on the 9-dimensional QUAK (loss) space. We observe QUAK significantly outperforms the latent space. 

From the results obtained with MNIST, we can make substantive conclusions about the design and future use of QUAK. First, we observe that high-dimensional QUAK spaces are highly effective provided a proxy signal prior is used that is similar to the potential anomalous signal. This proxy signal is used to reduce the dimensionality of the anomaly search. Despite using a proxy signal that differs from the true signal, we find that a linear discriminant trained on this proxy signal is sufficient to separate the individual digits 5 vs. 9. Furthermore, the proxy signal can be diluted within other incorrect signals and still give a comparable performance improvement. 

With our study on the MNIST dataset, we also observe that the latent space is not as effective as the N-dimensional QUAK (loss) space. We interpret this result as an indication that an individually trained VAE on each digit captures more information about a single-digit than a combined training intended to resolve all digits simultaneously. In the case where a VAE is sufficiently large and flexible to reproduce all digits as effectively as a VAE trained on a single digit, we expect that the use of the latent space in place of QUAK space should converge to the same level of performance.

\subsubsection{Concluding Observations}

Finally, we would like to conclude that our studies with MNIST illustrate that QUAK can be extended in several directions towards more complex approaches to search for anomalies. There is a rich and growing set of semi-supervised approaches that are being developed along similar lines. We believe that the addition of QUAK can help complement these other approaches and provides new insight towards the construction of semi-supervised algorithms to perform anomaly detection. 

As with the construction of QUAK space, we have effectively introduced a signal prior by separating digits or inserting signals. This will bias an anomaly search towards a specific region. However, we would like to stress that these added priors can deviate significantly from a true signal and still be an effective tool to search for anomalies. More generically, this concept lends itself to other types of physics analyses, such as measuring properties of a system where some, but not all, physical effects are known.

\subsection{LHC Olympics:}
Next, we perform an anomaly search with QUAK using the official LHC Olympics 2020 dataset\cite{kasieczka_gregor_2019_3596919}. The LHC Olympics 2020 consists of Pythia-simulated signal and background processes with detector smearing applied through the Delphes package\cite{deFavereau:2013fsa,Sj_strand_2015,Sjostrand:2006za, Sjostrand:2007gs}. A pure background sample consisting of QCD multijet(QCD) production using Pythia simulated dijet events is produced. This sample is used as a reference for simulated background events. 

A simulated two-prong signal sample is generated from a decay chain composed of a heavy charged spin-1 mediator $W^{\prime} \rightarrow XY,X\rightarrow\bar{q}q,Y\rightarrow\bar{q}q$, where $X$ and $Y$ are new physics mediators that decay to quark pairs and are light enough that their decay products fall within a single cone (i.e., a jet). Simulated three-prong signal events are generated from a heavy neutral spin-1 mediator decaying into a pair of heavy resonances that decay into three quarks. The intermediate resonances are also light enough that their decay products are reconstructed within single jets. A range of samples with different resonance masses and intermediary resonance masses are used to test various signals models' performance. 

Lastly, as a part of the LHC Olympics, a series of BlackBox datasets (1-3) are constructed. The BlackBox datasets' goal is to emulate a true data sample where a new physics signal is hidden within the background. In BlackBox 1, a sample with similar topology to the $W^{\prime}\rightarrow XY$ signal sample is hidden within a background sample. This sample is mixed into a QCD background sample using the same matrix element generation of the original Pythia sample, but with modified Pythia showering parameters so that the background does not match the pure background "simulation" sample. The variation in shower parameters is intended to emulate the observed deviation between QCD simulation and observed LHC multi-jet events. This means that data-driven techniques are needed within the BlackBox datasets to ensure the possibility of a signal. 

In the following study, we treat BlackBox 1 as a hypothetical dataset. We use the simulated signal and background samples to construct QUAK space and calibrate our approach. We then apply QUAK to BlackBox 1 and use it to search for anomalous features. Finally, we consider alternative anomalous signatures and discuss how QUAK compares to other anomaly searches. 

\begin{figure}[htbp]
\centering
\includegraphics[width=.8\linewidth]{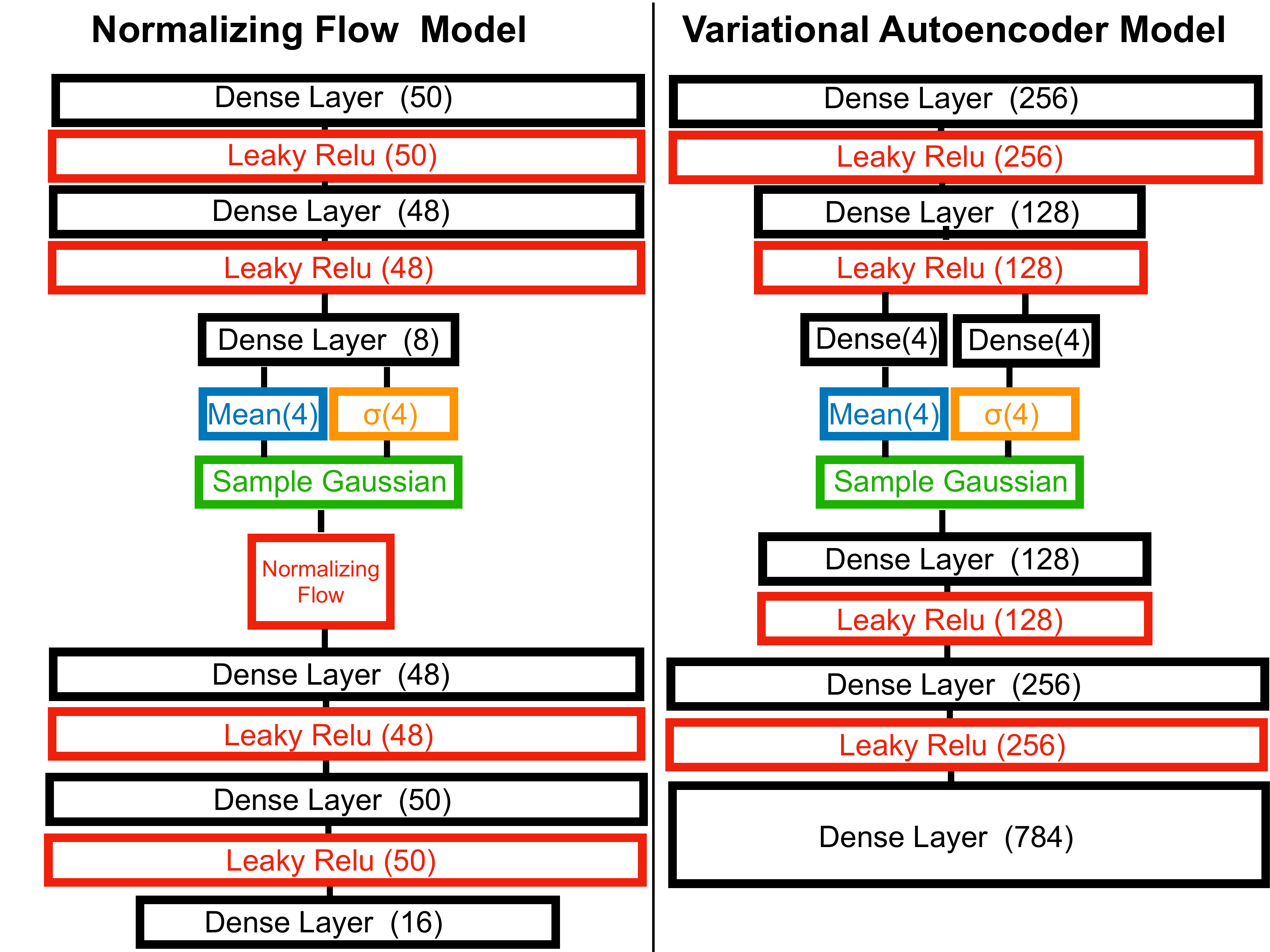}
\caption{ (Left) Illustration of the Normalizing Flow variational autoencoder network used for dijet identification. (Right) Illustration of the variational autoencoder used for the MNIST study. The number in parentheses indicates the number of output nodes of each layer. 
}
\label{fig:archs}
\end{figure}

To perform an analysis on the LHC Olympic dataset, we first reconstruct jets using the anti-$k_{t}$ algorithm with cone size parameter $\Delta R=0.8$~\cite{Cacciari:2005hq, Cacciari:2011ma}. From the reconstructed jets, we select the two highest energy jets. High-level jet features of each jet is computed and then these high level features are treated as inputs into the network for training. The high-level features consist of n-subjettines ratios ranging from 1-prong to 4-prong and the raw jet mass of the individual jets\cite{Thaler:2010tr,Thaler:2011gf,Thaler_2011,Datta_2017}. Training and testing are performed with 12 variables for each event (each jet: 4 n-subjettiness ratios, the total number of tracks, and the jet mass). 

To perform the unsupervised training and to construct QUAK space, an optimization scan of network architectures and parameters is performed on the simulated background and a single two-prong signal dataset. A broad range of architectures is considered, and the network architecture with the highest area under a curve separation (AUC) between the chosen signal and background is used. In this scan, we considered several different normalizing flow VAEs including planar flow, radial flow, non-volume preserving flow, and masked autoregressive flow. The optimized network we found was a Masked Autoregressive normalizing flows \cite{rezende2015variational,papamakarios2017masked}, with a latent $z_{dim}=4$ with 3 dense layers on either end. The details of the network are shown in Figure~\ref{fig:archs}. We apply a loss metric of mean-squared reconstruction error on each of the 12 variables with a KL-divergence term to regularize the sampling parameters for each training. 

As with other variational autoencoders, the KL-divergence term is added to constrain the samples mean and width of the Gaussian distributions to be near 0 and 1, respectively. This is added into the loss function with a tunable parameter $\beta$ that characterizes the relative scale of the cross-entropy auto-encoder loss with the KL-divergence term~\cite{Higgins2017betaVAELB}. We performed a hyperparameter scan for the $\beta$, and observed the optimal value of $\beta=10$,  and this value is used in the subsequent results. Finally, the QUAK space is constructed by computing the background loss and signal loss(es) with the KL-divergence term removed.

\subsubsection{Example search on LHC Olympics BlackBox 1}
\begin{figure}[htbp]
\centering
\includegraphics[width=.45\linewidth]{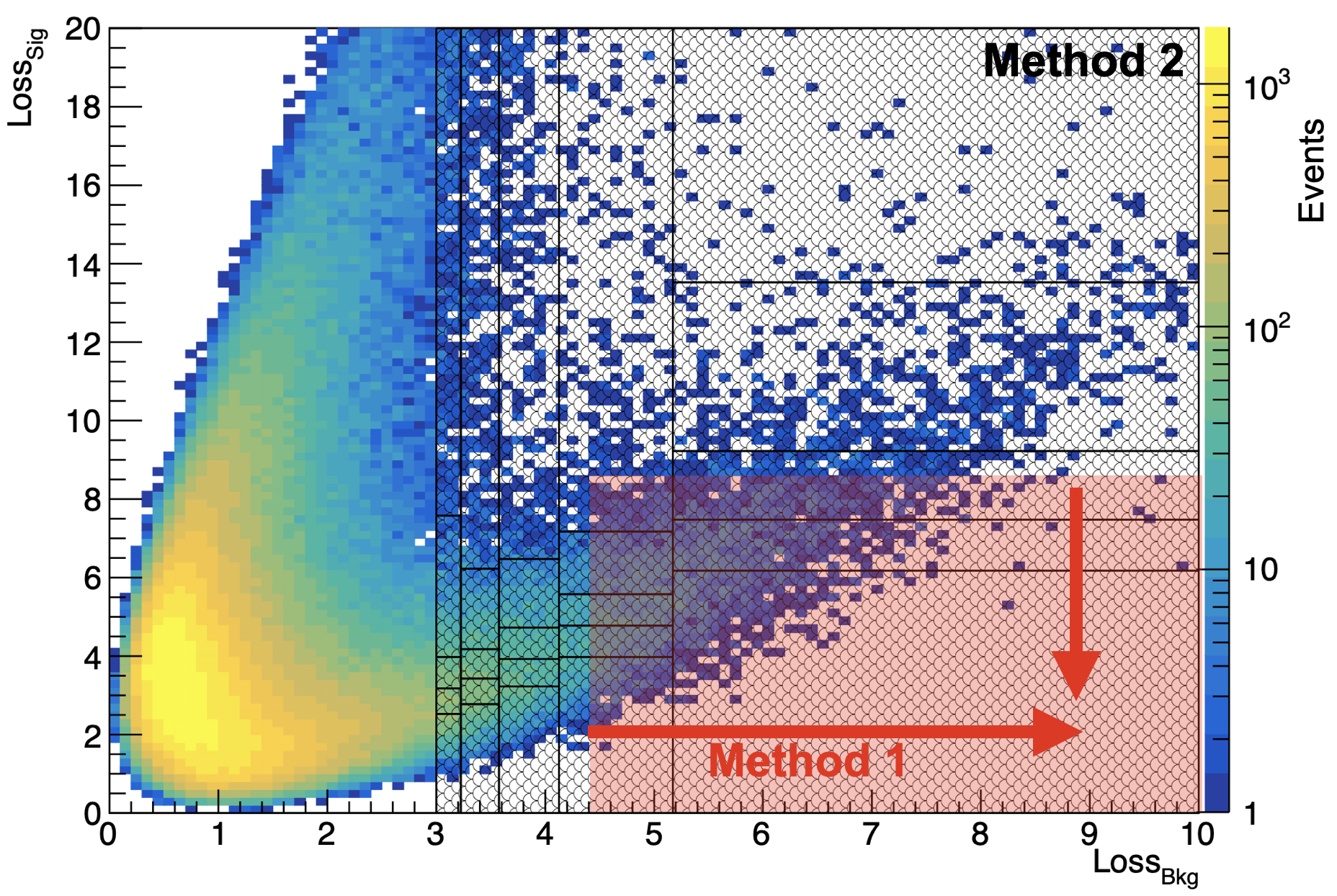}
\includegraphics[width=.45\linewidth]{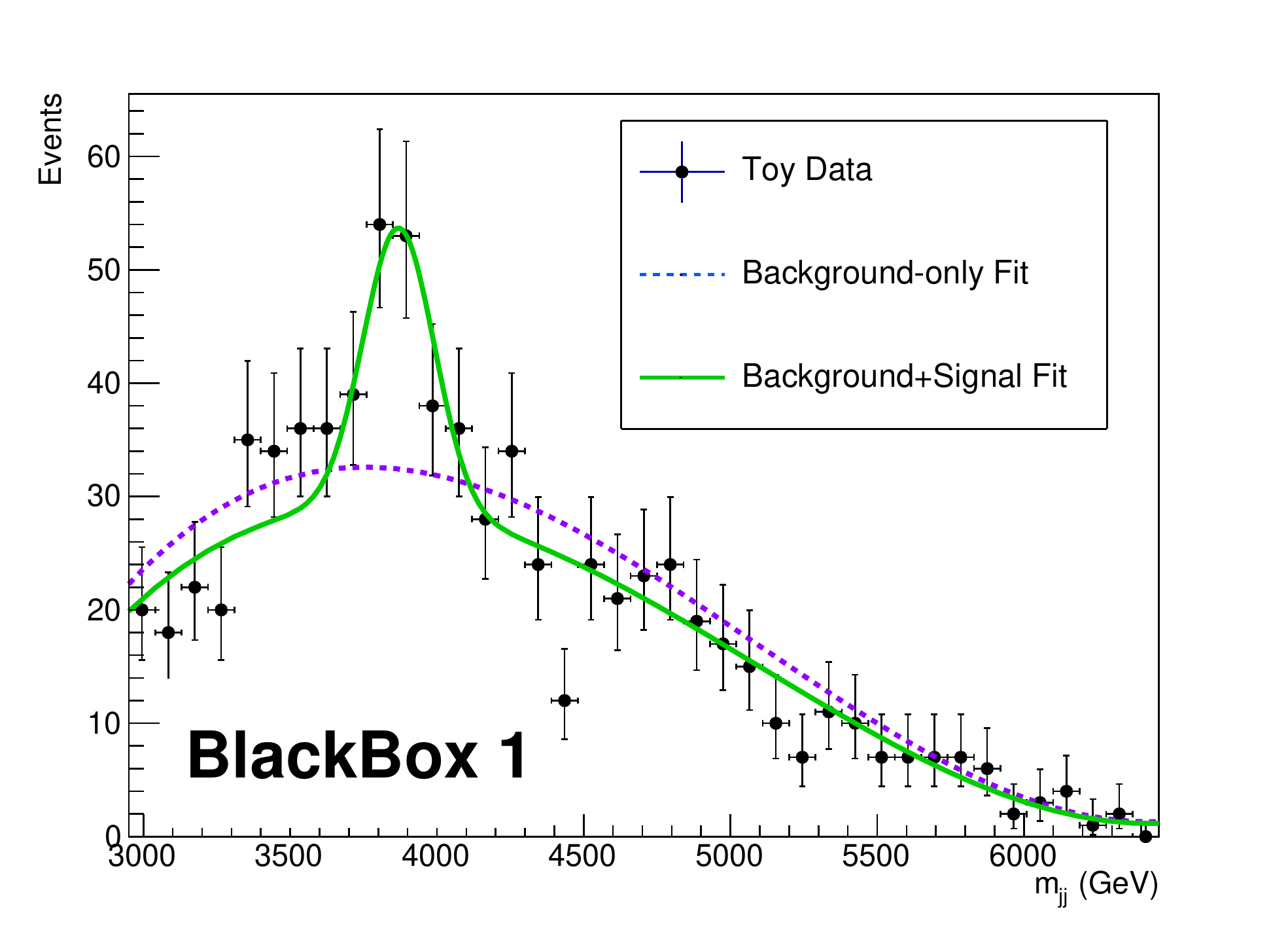}
\caption{ (Left) Illustration of the two methods used for the signal extraction. In Method 1, shown in red, we iteratively vary a selection on the signal loss $L_{Sig}$ and background loss $L_{bkg}$ and select regions of low signal loss and high background loss. In Method 2, we separate the events by the black shaded boxes shown corresponding roughly to a uniform populations of events within each shaded region. (Right) Dijet mass fit after performing an optimized selection of $L_{Sig} < 8$ and $L_{bkg} > 5.5$, a 3rd order Bernstein polynomial is used for the background while a Gaussian with a fixed width $\sigma/m_{jj}=0.03$ is used for the signal model. 
}
\label{fig:grid}
\end{figure}

First, we show the performance of QUAK on a search for a hidden signal in BlackBox 1 by constructing a 2-dimensional QUAK space constructed from loss on the simulated background sample with the loss of a single signal sample. For the signal loss, we use the loss trained on the "R\&D" signal dataset, which consists of a $W^{\prime}\rightarrow XY$ with the mass of the $W^{\prime}=3.5$~TeV, and the mass of $X=500$~GeV, and $Y=100$~GeV. For the rest of this section, we will refer to $L_{bkg}$ as the loss from the normalizing flow VAE trained on the background, and $L_{sig}$ as the loss from the normalizing flow background trained on the chosen signal. The 2D QUAK space applied to the BlackBox 1 dataset is shown on the left of Figure~\ref{fig:grid}. 
While it was not known at the time of the LHC Olympics, a secret signal is hidden within the BlackBox 1 dataset. This signal consists of roughly 900 signal events injected into a background of 1 million events. The injected signal consists of a 3.8~TeV $W^{\prime}$ resonance decaying to two, two-prong resonances, $X$, and $Y$. In this case, the mass of $X$ is 732~GeV, and the mass of the resonance $Y$ is 378~GeV. 

With the 2D QUAK space constructed, we consider two strategies to look for an excess within the space, method 1 and method 2. Both these strategies are shown in the left diagram of Figure~\ref{fig:grid}. In method 1, we systematically select events in a region of QUAK space and look for an excess of events. For this method, we require $L_{sig}$ to be small, and $L_{bkg}$ to be large. In method 2, we separate events within QUAK space into individual categories and perform a search for an anomalous resonance within each category separately. The individual categories are then combined, yielding a single anomaly search. For each category/selection within QUAK space, a fit of the dijet mass $m_{jj}$ is performed using a 3rd order Bernstein polynomial. The signal is assumed to be a Gaussian with a width of 3.0\% $\times m_{jj}$, roughly equal to the detector resolution of the simulated sample. To compute the significance, we perform a binned likelihood fit and compute the significance through the likelihood ratio using the asymptotic CLs method\cite{Cowan:2010js}. The p-values of individual categories are combined using Fisher's method. The significance is computed at a fixed dijet mass with a fixed width, and the mass is scanned over a range of 3~TeV $< m_{jj} <$ 6~TeV. 

\begin{figure}[htbp]
\centering
 \includegraphics[width=0.45\linewidth]{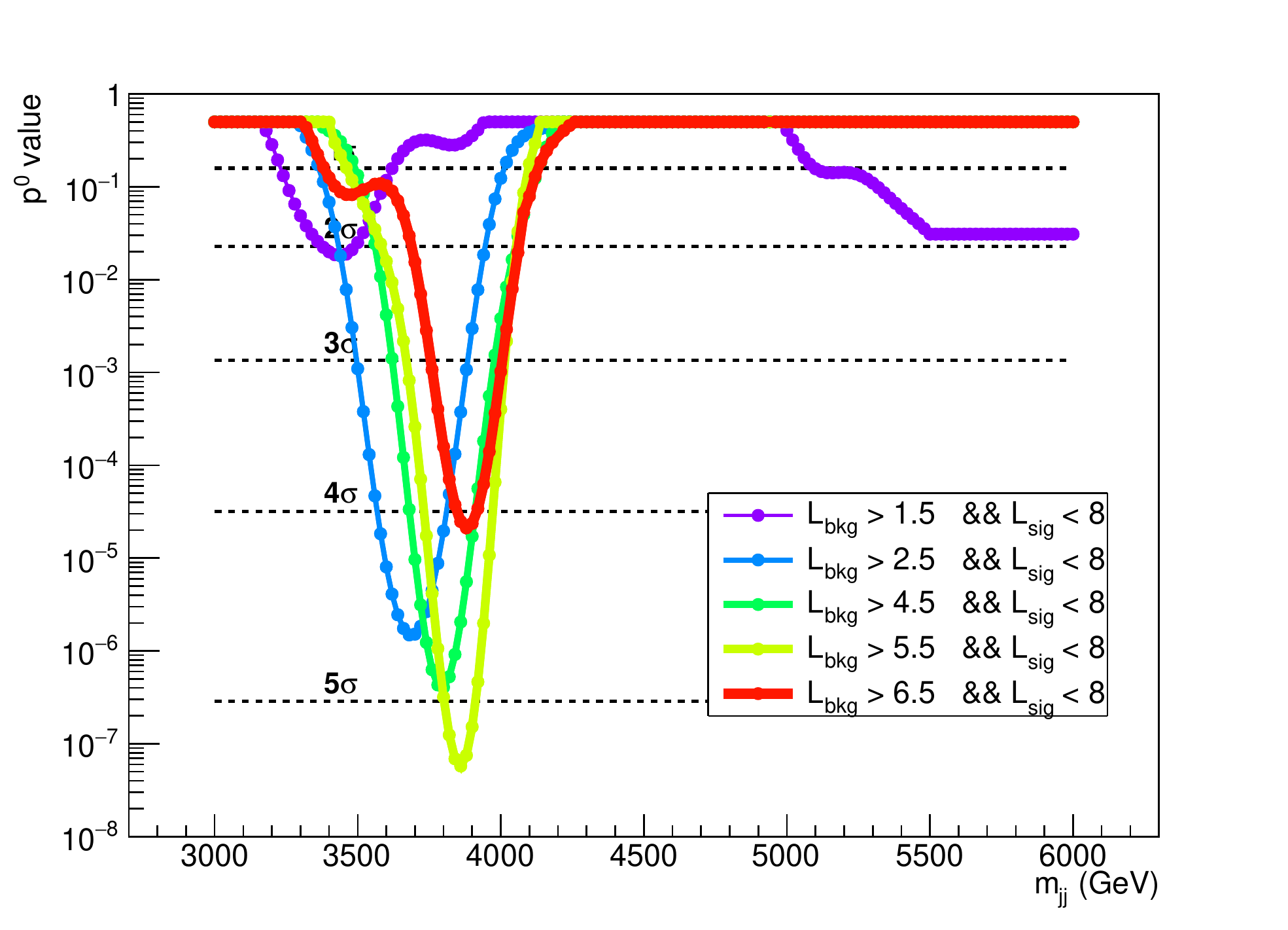}
 \includegraphics[width=0.45\linewidth]{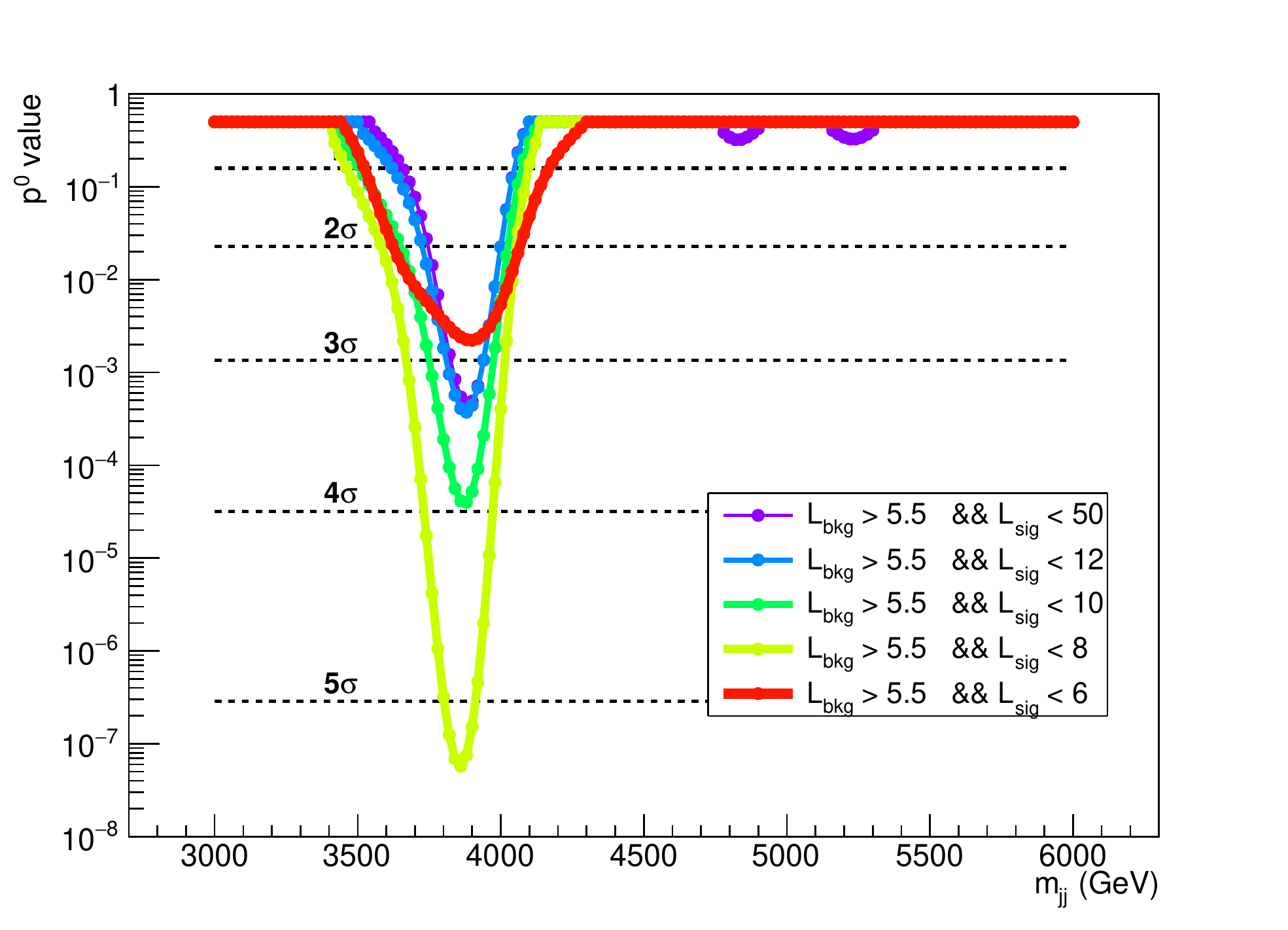}
\caption{Scan of the p-value significance in the dijet mass performed on a fit to the dijet mass distribution for (a) a tight selection on the signal loss $L_{sig}$ and varied selections on the background loss $L_{bkg}$, and for  (b) a tight selection on the background loss $L_{bkg}$ and a varied selection on signal loss $L_{sig}$. The significance is quoted is obtained from the asymptotic CLs method.}
\label{fig:method1}
\end{figure}

To find an excess with method one, we systematically vary the selection on the signal loss requiring $L_{sig} < X$, with $X$ varied. Additionally, we vary the selection, $Y$ on the background loss requiring $L_{bkg} > Y$. We then construct a grid that is evenly spaced over the 2D space and systematically fit the dijet mass distribution of the selected events. From this scan, we observe a maximum significance exceeding 5 standard deviations for a cut of $L_{bkg} > 5.5$ and $L_{sig} < 8$. The right of Figure~\ref{fig:grid} shows the resulting observed excess of events from the fit of the dijet mass after the optimized selection. In Figure~\ref{fig:method1}, we show the change in significance when we vary both the selection on the background loss and the signal loss. We observe that a selection on both signal and background loss variables is needed to obtain the large observed significance. 

While method 1 is illustrative to find an excess, the quoted significance cannot be obtained from this approach since it is inherently biased towards finding large excesses. This results from the fact that the algorithm self-selects an excess in data by scanning variables and looking for an excess, whether it is real or not. In place of quoting the significance from method 1, we instead perform method 2. With method 2, we apply a preselection cut on the background loss, $L_{bkg} > 3$, roughly corresponding to a 2 percent background probability. After that preselection, we construct N uniformly populated categories by first dividing $L_{bkg}$ into evenly populated regions and then within each region dividing along $L_{sig}$ into evenly populated regions. The different regions used in this method are labeled as method 2 in Figure~\ref{fig:grid}. With these uniformly populated bins, we perform a separate fit to the dijet mass distribution in each category. The uniform number of events allows for the same order Bernstein polynomial(3rd) to be used for every category of the fit. Additionally, we assume no correlation between fits. The categories are then combined into a single quoted p-value. To assess the performance of this method, we consider a 3x3,4x4,5x5 binned search in $L_{bkg}$x$L_{sig}$ space. As a check of this method, we applied this method to a simulated sample with no signal injected. We found one excess above 2 standard deviations, consistent with the expected number of deviations expected from a random sampling. 

\begin{figure}[htbp]
\centering
\includegraphics[width=.45\linewidth]{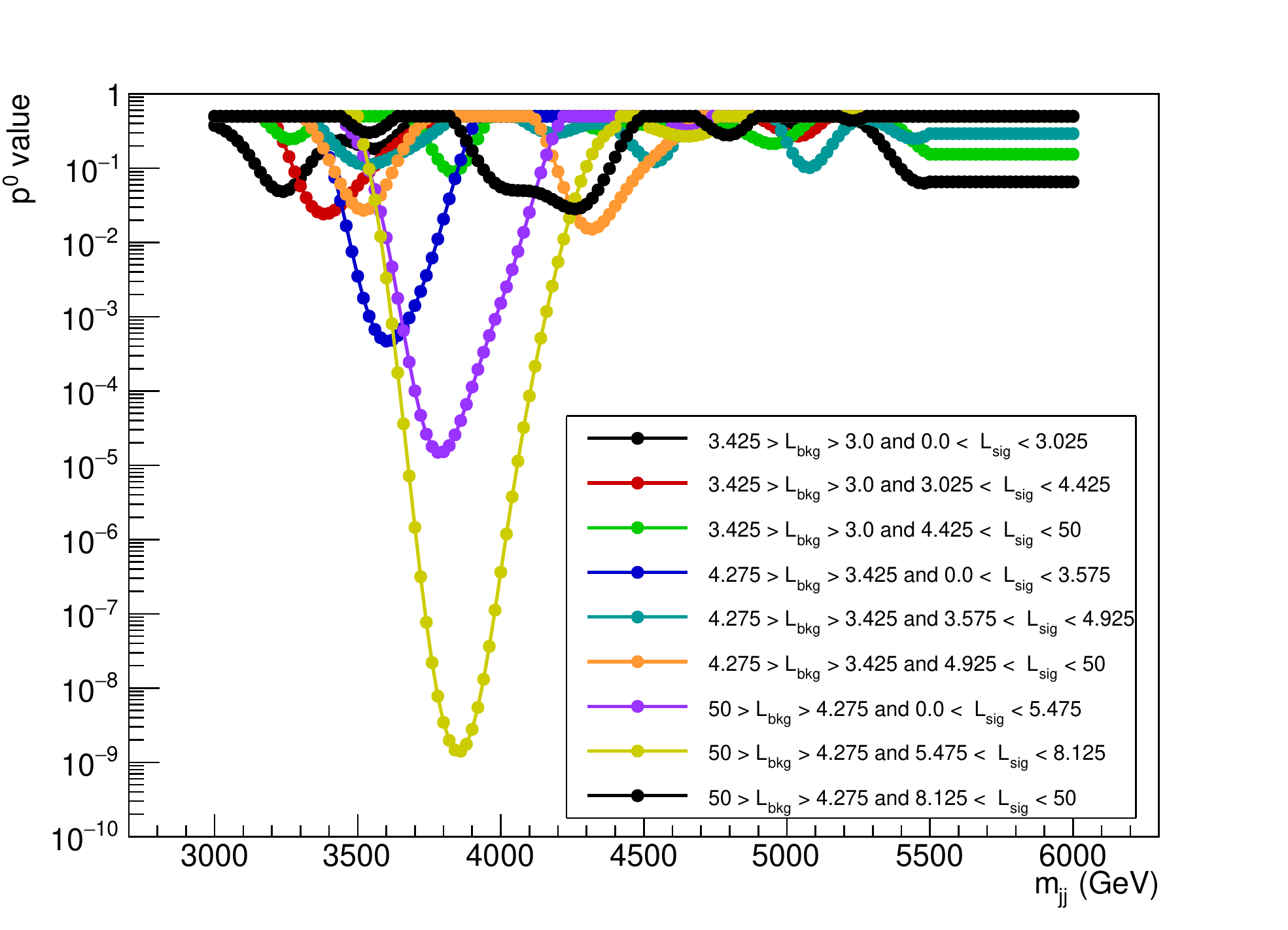}
\includegraphics[width=.45\linewidth]{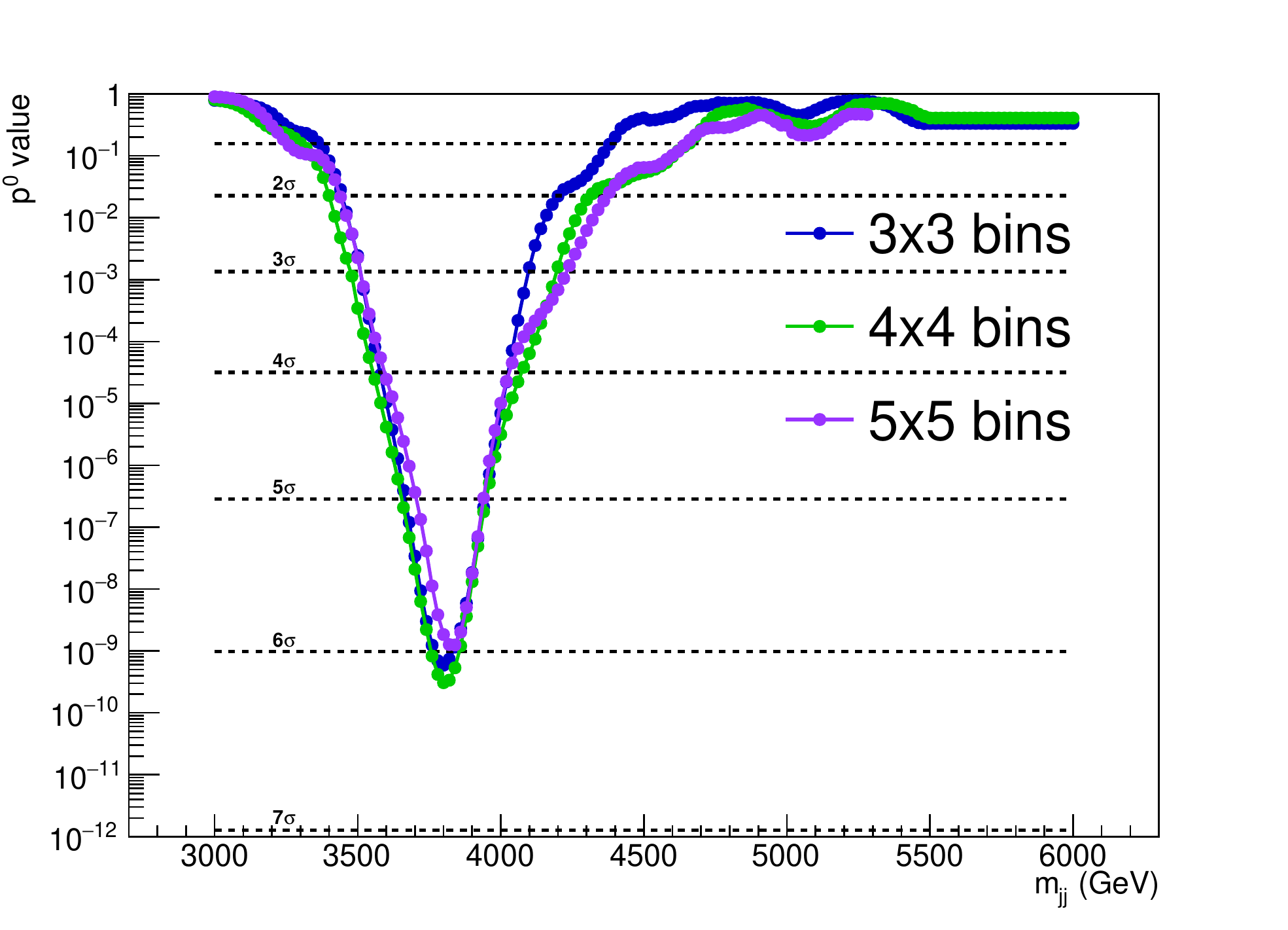}
\caption{ Scan of the significance of an excess quoted in the p-value significance of a Gaussian resonance on top of a fitted background using the asymptotic CLs method for (Left) individual categories in $L_{Sig}$ and $L_{Bkg}$, and (Right) combined categories using a uniform populated set of 9 bins (3 in $L_{Sig}$, 3 in $L_{Bkg}$) 16 bins (4,4), and 25 bins (5,5). 
}
\label{fig:method2}
\end{figure}

Figure~\ref{fig:method2} shows the individual p-values obtained from method 2 for 9 separate bins; one region clearly dominates the p-value computation. The combined results yield an excess behyond 6 standard deviations. Additionally, we observe a consistent p-value profile when we perform the anomaly search in 9,16 and 25 bins. This approach effectively minimizes the look-elsewhere effect and allows us to quote a significance measurement within a broad, model independent, selected region. 

In summary, we find an excess of events with a significant deviation from the background of over five standard deviations. The excess is found to be at 3.8~TeV, consistent with the hidden signal injected within the dataset. The excess of events is located in the region of low signal loss and high background loss, consistent with an anomalous signature similar to the test signal used to construct $L_{sig}$.

\subsection{Performance of QUAK on different signals}
Going beyond the example anomaly search performed with the BlackBox 1 dataset, in the following section, we consider the performance of QUAK on different signals and with higher dimensional QUAK space. Furthermore, to understand how effective QUAK is at finding new physics signatures, we characterize the performance of QUAK against fully supervised networks.

\begin{figure}[htbp]
\centering
 \includegraphics[width=0.45\linewidth]{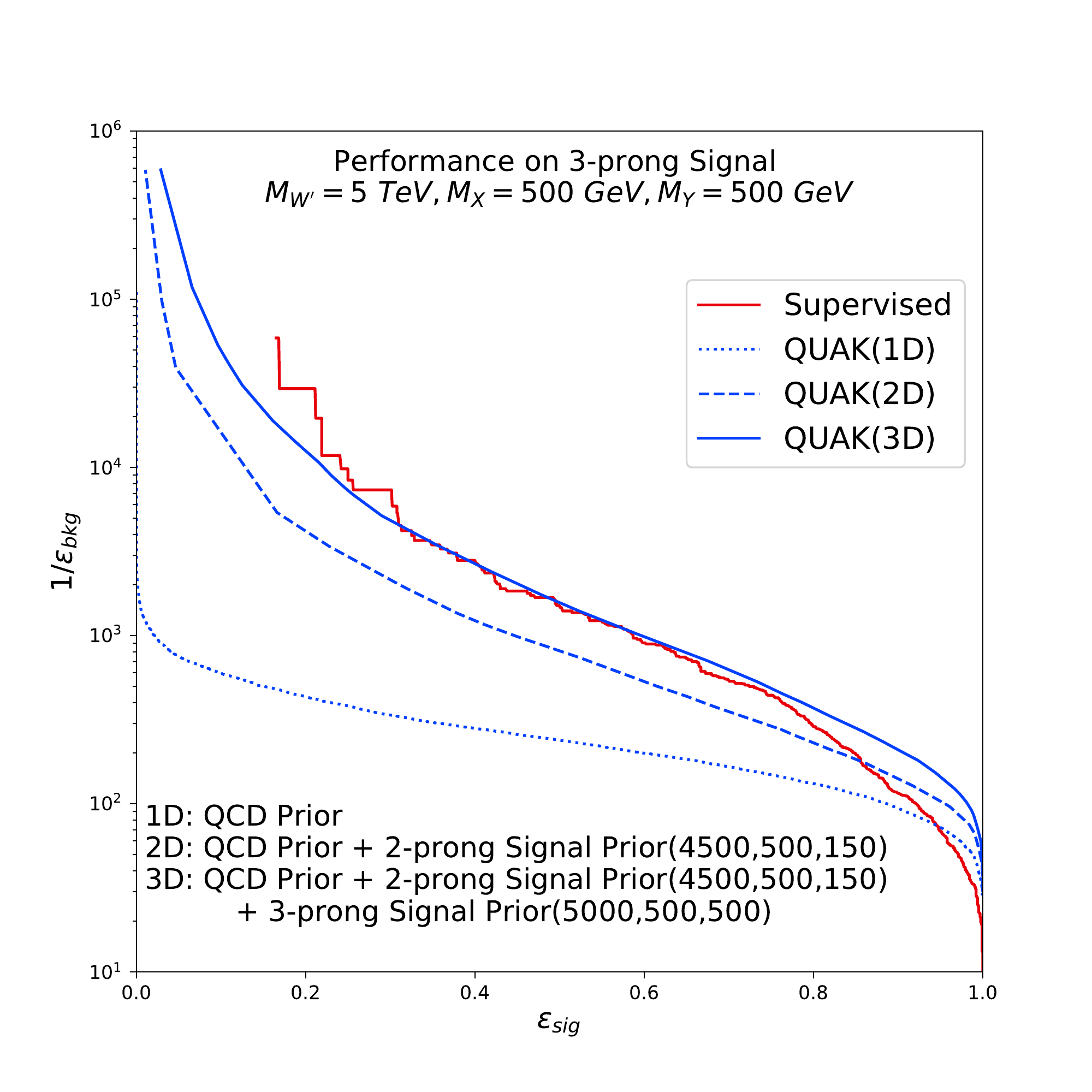}
 \includegraphics[width=0.45\linewidth]{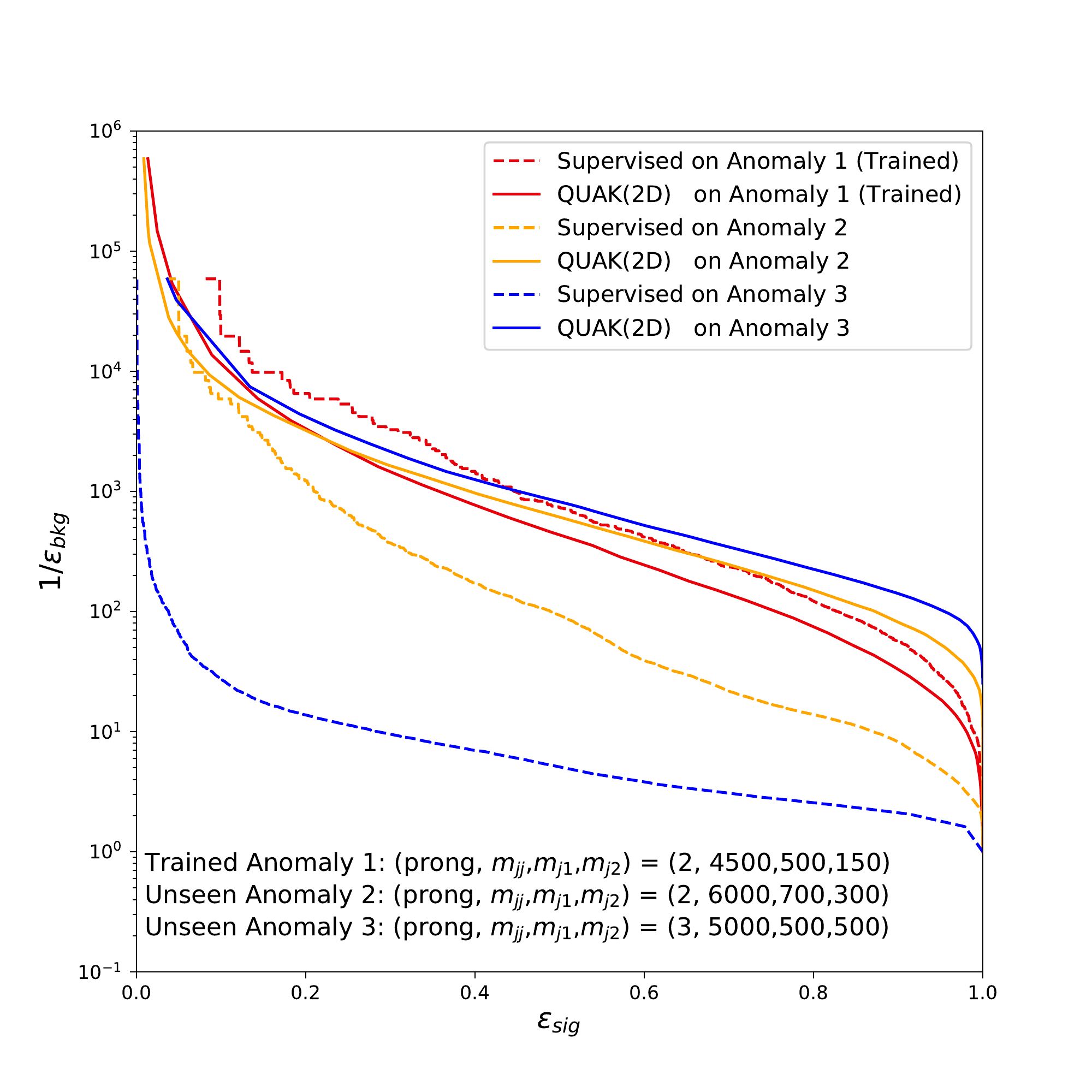}
\caption{(Left) Receiver Operator Characteristic (ROC) for a resonant signal compared to QCD background; the resonant signal has mass $m_{jj}=5$~TeV~ that decays into two three-pronged objects with mass $m_{j1}=m_{j2}=500$~GeV~ against a background. Signal priors are labeled in the legend by $(m_{jj},m_{j1},m_{j2})$. Performances are shown in blue for the 1D QUAK (QCD prior only), 2D QUAK (QCD prior and a 2-prong signal prior) , and 3D QUAK (+ 3-prong signal prior), and shown in red for a fully supervised training on the true signal prior against background. Jet masses $(m_{j1}, m_{j2})$ are excluded in the training of the supervised classifier to mitigate model dependence. (Right) ROC for signal versus QCD background computed from a selection on \emph{two} neural networks consisting of: QUAK(2D) (solid) for one network and a  supervised network (dashed) for the other network. Both networks are trained once and constructed using the same 2-prong signal prior.  The networks are then applied to different signal models. For both QUAK and the supervised network a signal prior of a resonance $W^{\prime}\rightarrow XY$ with $X$ and $Y$ decaying to pairs of quarks having masses $(4500,500,150)$ is used in the training.}
\label{fig:rocs}
\end{figure}

Figure~\ref{fig:rocs} shows the performance of different QUAK methods in separating the background from a resonance decaying to two 3 prong resonances $Z^{\prime}\rightarrow \bar{t}^{\prime}{t}^{\prime}$ with masses $m_{Z^{\prime}}=5$~TeV, and $m_{t^{\prime}}=m_{\bar{t}^{\prime}}=0.5$~TeV. To understand the gain in adding signal loss to search for anomalies, we first consider anomaly detection performance with a single normalizing flow autoencoder trained on the background sample. In the following comparisons, we will label the single autoencoder as 1D QUAK since it is just a 1-dimensional QUAK space on $L_{bkg}$. To see the gain in using signal loss, we then construct 2D QUAK, which consists of 2D space with one axis being $L_{bkg}$, and the other axis $L_{sig}$. For $L_{sig}$, we choose as a signal prior the loss from a training on $W^{\prime}\rightarrow XY$ with resonances $X$ and $Y$ decaying to two prongs , and the masses $m_{W^{\prime}}=4.5$~TeV, $m_{X}=500$~GeV, and $m_{Y}=150$~GeV. Finally, we construct a 3D QUAK space by appending a 3rd loss $L_{sig2}$ to the 2D QUAK space. Here $L_{sig2}$ is computed from a network trained on the same $Z^{\prime}$ signal used for comparison. 

To compare the performances in Figure~\ref{fig:rocs}, Receiver Operator Characteristics (ROCs) are obtained by systematically scanning the 2D or 3D QUAK spaces linearly in each dimension of QUAK and requiring events to be selected from the region of minimum signal loss, $L_{sig_{i}} < X$, and maximum QCD loss, $L_{bkg} > Y$. This is done for the background sample and the signal sample; the resulting background and signal efficiencies for each selection are presented on the plot. The ROC characterizes the discrimination power for a systematic n-dimensional search to find anomalies. 

The performance comparison when going from 1D to 2D to 3D QUAK is shown on the left in Figure~\ref{fig:rocs}. In addition to QUAK, we compare QUAK to a supervised classifier trained on the same inputs, excluding the jet masses. The chosen signal prior to the supervised training is the same signal used in the ROC computation. The fully supervised network consists of a 4 layer multi-layer perceptron with batch normalization ~\cite{ioffe2015batch} and dropout\cite{hinton2012improving}. Jet masses are excluded from the training since they have the ability to isolate specific kinematic features present in the samples that are not available in a normalizing flow autoencoder, which, by construction, is not strongly correlated with jet masses. Furthermore, in a realistic analysis scenario, similar supervised searches aim to decorrelate discrimination against the mass so that fitting and template based background methods can be used to extract the signal\cite{Dolen:2016kst,Moult:2017okx, Stevens:2013dya, Shimmin:2017mfk,Bradshaw_2020, ATL-PHYS-PUB-2018-014, Kasieczka:2020yyl,Kasieczka:2020pil,Sirunyan:2019jbg, Sirunyan:2020hwz, Sirunyan:2017nvi, Sirunyan:2019vxa, Aad:2019fbh, Sirunyan:2019vgj, Sirunyan:2018rlj, Aad:2019uoz, Sirunyan:2017dgc, Sirunyan:2017jix}. 

By comparing 1D, 2D, and 3D QUAK, we observe an increase in the search's sensitivity by adding more approximate signal priors. The addition of the approximate priors approaches, and in some places exceeds, the performance of a supervised discriminator computed by training the same inputs on the known signal. Interestingly, much of the gain in the separation of the 3-prong signal arises by adding the 2-prong signal prior despite these signals having manifestly different topologies. As a reference to existing studies, previously proposed single autoencoder searches would comprise 1D QUAK, so we see a large improvement over previous single autoencoder based searches given this modified search strategy. 

With QUAK, we conclude that even if the signal priors are not accurate, we gain a sizable performance improvement. We theorize that the added information present in the signal loss helps isolate "signal-like" anomalies from other anomalous features present within the background. Through the construction of the QUAK space, we also demonstrate that incorrect signal priors, whether they result from inaccurate simulation or signal model choice, can still be a powerful discriminant when searching for new physics. 


In the right part of Figure~\ref{fig:rocs}, we contrast QUAK with a conventional new physics approach based on supervised learning. We train a supervised classifier to discriminate background against a specific two-pronged signal. We then apply this supervised classifier to a range of different signal models. The use of a supervised classifier is reflective of current supervised based machine learning searches used at the LHC since there is no guarantee that a signal model's simulation is consistent with a true signal. For the supervised training, we choose a $W^{\prime}\rightarrow XY$ with resonances $X$ and $Y$ decaying to two prongs, and the masses $m_{W^{\prime}}=4.5$~TeV, $m_{X}=500$~GeV, and $m_{Y}=150$~GeV. A fully-connected network is used for the supervised classifier (4 hidden layers with batch normalization ~\cite{ioffe2015batch} and dropout\cite{hinton2012improving}). Both QUAK and the supervised classifier are trained on the same raw inputs, signal prior, and background prior (a QCD prior and a 2-prong prior)~\cite{ioffe2015batch,hinton2012improving}, again exluding jet masses for the supervised network. 

With the supervised network, we observe a general trend where the supervised classifier performs worse that QUAK as the signal deviates further from the chosen 2-prong prior used to train the supervised classifier. With the 3-prong signal, we find extremely poor performance with the supervised classifier. With QUAK, we observe a relatively stable separation between background and signal even as the test signal further deviates from the chosen signal prior. We conclude that QUAK incorporates signal priors in a more efficient way than supervised classifiers, and by using QUAK, we can do a more efficient scan of the possible space of new physics. For searches where the signal prior is partially known (to within uncertainties), QUAK has the potential to mitigate loss in sensitivity since deviations within this space have a smaller impact on the separation power when compared with supervised models. 

We stress that the assumption that the signal is "approximately" correct is an assumption that is implicit in every new physics search at the LHC. All physics searches at LHC can be reduced to separating hypothetical signals from the background with a signal represented by a simulation that is our closest guess to what we believe the true signal is. With QUAK, we have constructed a space that attempts to mollify the differences between incorrect assumptions. 

QUAK is a likelihood based approach, which aims to describe all of the variables of a specific process. As a consequence, our approach captures full distributions of the data. Whereas, supervised networks aim to optimize a decision boundary between a hypothetical signal, and a background under the assumption that a signal and background simulation are accurate representations. Methods like QUAK, which are sensitive to the full likelihood will be impacted less by deviations in simulated parameters from truth when compared with a supervised network. As a consequence, based on our observed performance, we believe  QUAK and other likelihood based methods will be more robust to systematic variations, and uncertainties within the simulated signal and background modelling. 

Minimizing the impact of systematic uncertainties on network design has been the focus of many studies. Previous approaches have used adversarial neural networks and other bounding principles to constrain the impact of systematic uncertainties on a network performance\cite{Wunsch:2019qbo, Englert:2018cfo,Xia:2018kgd,Louppe:2016ylz}. Based on our observed variations, the use of QUAK space in conjunction with these approaches can lead to further improvements in this area. In future LHC analyses, we expect that QUAK and other likelihood based approaches will be able to outperform supervised networks by limiting their sensitivity to inadequate simulations.	

\subsubsection{Concluding Observations}
In summary, we find that QUAK space's construction can be used to find anomalies over a broad range of signatures. It is robust against variations in signal choice. Additionally, incorrect signal priors within the QUAK space construction can still significantly enhance the detection of anomalies. Furthermore, QUAK is not strongly sensitive to model variations or incorrect signal choices, allowing for many new physics models to be probed by one search. Finally, we find that the QUAK construction can approach and sometimes exceed supervised networks.

%% file: conclusion.tex
In summary, we propose the exploration of a new algorithm, QUAK, to perform model independent searches. We demonstrate this work in the context of new physics search at the LHC. We observe that the addition of approximate priors to anomaly loss allows for enhanced identification of anomalies by providing generic "signal-like" or "background-like" features to help in identification. With QUAK, we have presented an approach that effectively adds these priors without degrading the sensitivity of a prior-free model. Furthermore, by relying on unsupervised learning techniques, we have allowed the neural networks to self-organize leading to a network performance that is robust against large variations.  QUAK is broadly applicable to many different problems and can improve both anomaly searches and searches where large uncertainties are present on the signal modeling. 

In particular, we have demonstrated QUAK as a new approach to construct anomaly searches using autoencoders and injecting signal priors. We demonstrate that even with incorrect priors, we can enhance anomaly searches by creating an anomalous space that is more conducive to searching for new physics models. Our work demonstrates a significant improvement in discrimination power on new physics models, when compared with previous single autoencoder approaches, even when the true signal deviates significantly from the assumed signal prior. 

Furthermore, we observe that QUAK can approach or even exceed the sensitivity of supervised searches. Recently, a number of studies have observed that unsupervised and semi-supervised approaches are capable of exceeding the performance of direct supervised training. This has been attributed to the self assembly of the weights that occurs when training an unsupervised network. Lastly, while we have not directly compared this method to the classification without labels approach to anomaly searches, we would like to highlight that these approaches are mutually exclusive, and they can be used in concert. 

We first characterized QUAK approach through an example using the MNIST dataset and a variational autoencoder architecture. We observe that QUAK can be extended to solve higher dimensional problems, provided a scheme to choose signal priors. Additionally, we observe that the QUAK construction is found to be more effective than a similar approach relying on the latent space. 

To demonstrate the use of QUAK for anomalous features in LHC-like dijet data, we constructed QUAK space by using normalizing flow variational autoencoders trained on the n-subjettiness jet observable inputs. While we found both normalizing flows and n-subjettiness observables are effective for this approach; we would like to stress that this method can be used with other types of deep learning methods, and observable calculations. In particular, we think that recent work with energy flow polynomials, and earth movers distance can be used within QUAK in the construction of the space and training, respectively, to build a neural network with characteristic physical features \cite{Komiske_2018,Komiske_2019}. We see this as an exciting avenue for future study.

Based on the small variation across signal models, we observe that QUAK lends itself well to measurements, not just anomaly searches, where the data is high-dimensional, and where the signal model is not well known. This is true for many fields in physics where simulation is limited, including gravitational wave modeling, identifying astrophysical signatures, and quark-gluon plasma physics. With QUAK, we can naturally construct signal models that carry some of the signal features, allowing for enhanced anomaly identification under precepts of physical principles. This can potentially be used in high energy physics to perform measurements of new particle decays from existing known resonances, such as the Higgs Boson\cite{harris2019approach}. Outside of physics, this work builds on recent results that relate to semi-supervised anomaly detection and semi-supervision itself. We see this work as an initial step towards a broad range of new deep learning approaches that can have a significant impact in many new fields.

%% file: acknowledgements.tex
\section*{Acknowledgments}
P.~H., D.~R., M.~Y. are partially supported by NSF grants \#1934700,  \#1931469, and the IRIS-HEP grant \#1836650.
Additionally we would like to think NSF Institute for AI and Fundamental Interactions (Cooperative Agreement PHY-2019786).
We would like to thank Nhan Tran, Cristina Mantilla Suarez, Jesse Thaler, and Javier Duarte for useful comments. We thank Erik Katsavounidis, Tri Nguyen, and Alec Gunny for interesting discussions. Additionally, we would like to thank David Shih, Gregor Kasieczka, and Ben Nachman for their help with the LHC Olympics dataset.